\documentclass[twocolumn,twoside,slac_two]{revtex4}
\usepackage{graphicx}
\usepackage{fancyhdr}
\begin{document}

\title{Radiation Mechanism of the Soft Gamma-ray Pulsar PSR B1509-58}

%

\author{Y. Wang, J. Takata $\&$ K.S. Cheng}
\affiliation{Department of Physics, University of Hong Kong, Pokfulam Road, Hong Kong}

\begin{abstract}
The outer gap model is used here to explain the spectrum and the energy dependent light curves of the X-ray and soft $\gamma$-ray radiations of the spin-down powered pulsar PSR B1509-58.
 In the outer gap model, most pairs inside the gap are created around the null charge surface and the gap's electric field separates the two charges to move in opposite directions. 
 Consequently, the region from the null charge surface to the light cylinder is dominated by the outflow of particles and that from the null charge surface to the star is dominated by the inflow of particles. 
 The inflow and outflow of particles move along the magnetic field lines and emit curvature photons, and the incoming curvature photons are converted to pairs by the strong magnetic field of the star. These pairs
  emit synchrotron photons.
 We suggest that the X-rays and soft $\gamma$-rays of PSR B1509-58 result from the synchrotron radiation of these pairs,
 and the viewing angle of PSR B1509-58 only receives the inflow radiation.
The magnetic pair creation requires a large pitch angle, which makes the pulse profile of the synchrotron radiation distinct from that of the curvature radiation. 
We carefully trace the pulse profiles of the synchrotron radiation with different pitch angles. We find that the differences between the light curves of different energy bands are due to the different
 pitch angles of the secondary pairs, and the second peak appearing at $E>10$MeV comes from the region near the star, where the stronger magnetic field allows the pair creation to happen with a smaller pitch angle.

\end{abstract}

\maketitle

\thispagestyle{fancy}


\section{INTRODUCTION}
PSR B1509-58 (hereafter PSR B1509) is a unique subject of the multi-wavelength study of the high energy radiation mechanism of the spin-down powered pulsar. 
It is a young (about 1600 years old) spin-down powered pulsar with a period of about 150 ms and a surface magnetic field of $\sim 1.5\times{}10^{13}$G, which is much stronger 
than those of most of the canonical pulsars ($\sim{}10^{12}$G). Multi-wavelength observations show that, from 0.1 keV to 10 MeV, 
  PSR B1509 has a wide single peak \cite{Beppo_Obs, OSSE_Obs, PCA_Obs, COM_Obs, RXTE_long}, whose shape can be described by two Gaussian functions, 
  when $E>10$MeV, another peak shows up near the peak of radio emission \cite{COM_Obs, AG_Obs, Fermi_Obs}.

PSR B1509 is unique for its soft $\gamma$-ray spectrum and a very strong surface magnetic field among the known $\gamma$-ray spin-down powered pulsars. 
It is suggested that there may be a new class of spin-down powered pulsar with high magnetic field ($>$10$^{13}$G) and soft $\gamma$-ray radiation, called soft $\gamma$-ray pulsar \cite{SGP}. 
And there is at least another pulsar that is similar to PSR B1509, PSR J1846-0258, a young pulsar with $P \sim 324$ms and surface magnetic field $B_{s} \sim 4.9\times{}10^{13}$G, 
 which cannot be detected by Fermi but by X-ray telescopes \cite{J1846, J1119}. Therefore, a multi-wavelength study of the radiation mechanism of the PSR B1509 
 can be a bridgehead to investigate this possible new class of pulsars.

Recent observations show that, two kinds of pulsars: the spin-down powered pulsar with $B_s\sim{}10^{12}$G and the magnetar with $B_s\sim{}10^{14}-10^{15}$G, have no clear boundary
 between them. It is believed that the radiation of a magnetar (AXP or SGR) is powered by the energy stored in the super strong magnetic field with $B_s>10^{14}$G \cite{Magnetar}.
 However, the discoveries of the low magnetic field soft $\gamma$-ray repeater \cite{0418_obs, 1822_obs}, the radio emission from magnetar \cite{1810_obs, 1547_obs, 1622_obs}
 and the magnetar-like outburst of
 spin-down powered pulsar with high magnetic field \cite{J1846} challenge the existing theoretical models. The study of the radiation mechanism of the very high magnetic
      field spin-down powered pulsar, and the connection between the canonical spin-down powered pulsar and magnetar, can help us to understand the physics of the two groups of neutron stars. As a family
       member of the high magnetic field spin-down powered pulsars, PSR B1509 may provide some insights for understanding the differences between typical spin-down powered pulsars and magnetars.

Here we give explanations for the spectrum and the energy dependent light curves of the X-ray and soft $\gamma$-ray radiations of the spin-down powered pulsar PSR B1509-58, based on the outer gap model \cite{CHR}.
Cheng, Ruderman \& Zhang \cite{CRZ00} showed that most electron/positron pairs created inside the outer gap are produced near the null charge surface; the strong electric field inside the gap
 separates the opposite charges to move in opposite directions. Therefore from the null charge surface to the light cylinder the radiation is mainly outward, and from the null charge surface 
 to the star the radiation is mainly inward. Since the light cylinder of PSR B1509 is much bigger than that of the Crab pulsar, most outflow curvature photons can escape from photon-photon pair 
 creation. If the line of sight is in the direction of outgoing radiation beam, the spectrum of PSR B1509 should be a characteristic pulsar spectrum, namely a power law with exponential cut-off 
 spectrum as predicted by Wang, Takata \& Cheng \cite{WTC10}. Obviously this conflicts with the observation. Therefore, we propose that the viewing angle of PSR B1509 is in the
  direction of incoming radiation beam, where the curvature photons of hundreds MeV are converted into pairs by the strong magnetic field. The observed X-rays and soft $\gamma$-rays are synchrotron
   radiation of these pairs. 
   
\section{SIMULATIONS OF THE SPECTRUM AND THE ENERGY DEPENDENT LIGHT CURVES}

\begin{figure}[t]
\centering
\includegraphics[width=80mm]{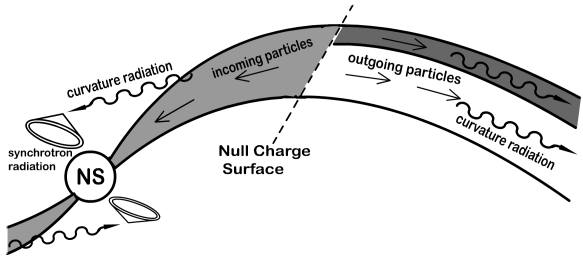}
\caption{The sketch of the structure of the gap.}
\label{struct}
\end{figure}

Figure~\ref{struct} shows the structure of the acceleration region. Outside the null charge surface, the structure of the gap can be simplified as a two-layer structure \cite{WTC10, WTC11}. 
From the null charge surface to the stellar surface, the accelerating electric field is screened out. Outside the null charge surface, the particle can get the equilibrium state
that the particle loses the energy obtained from the potential drop via curvature radiation. Hence its Lorentz factor can be obtained from  $eE_{\parallel}(\vec{r})c=2e^2c\gamma^4(\vec{r})/3s^2(\vec{r})$.
On the other hand, for the particle moving from the null charge surface to the star, the energy loss by the curvature radiation cannot be compensated for by the reduced electric field. Therefore, we solve numerically the
 evolution of the Lorentz factor, which is given by
\begin{equation}
mc^2\frac{d\gamma}{dt}=-\frac{2}{3}\gamma^4(\vec{r})\frac{e^2c}{s^2(\vec{r})}+eE_{\parallel}(\vec{r})c.
\end{equation}  

We trace the field lines in the gap from the stellar surface to the light cylinder.
At each step $\vec{r}$ in the field line, we calculate the pulse phase and viewing angle $(\psi, \zeta)$ of the curvature radiation, whose direction is tengential to the field line in the co-rotating frame.
Then we trace the direction of each incoming curvature radiation to find the place where the magnetic pair creation happens. The synchroton radiation of the pair is a hollow cone, where the $(\psi, \zeta)$ of each direction is calculated. 

\begin{figure}
\centering
\includegraphics[width=80mm]{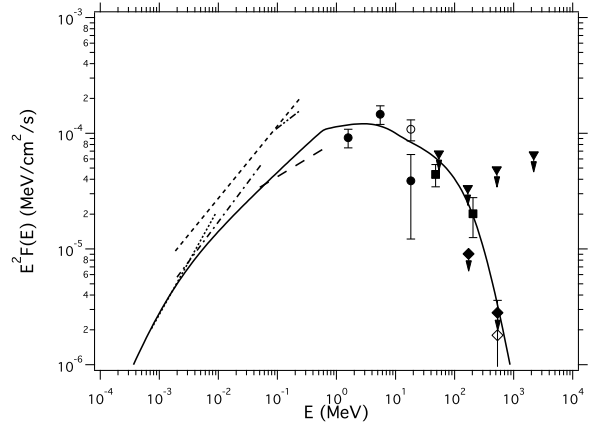}
\caption{The simulated phase averaged spectrum of PSR B1509 (solid line), comparing with the observed data provided by EGERT (triangles), COMPTEL (circles \cite{COM_Obs}),
AGILE (filled squares \cite{AG_Obs}), Fermi (diamonds \cite{Fermi_Obs}), ASCA (dot line \cite{ASCA_Obs}), Ginga (dot-dashed line \cite{Ginga_Obs}), OSSE (long-dashed line \cite{OSSE_Obs}), 
Welcome (dot-dot-dashed line \cite{Wel_Obs}), and RXTE (short-dashed line \cite{PCA_Obs}).} 
\label{Spectrum}
\end{figure}

\begin{figure}[t]
\centering
\includegraphics[width=80mm]{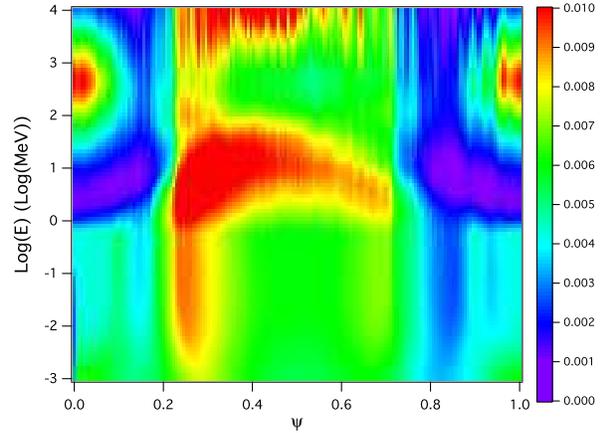}
\caption{The simulated energy dependent light curves in the pulse phase and energy plane. 
The color represents the percentage of the number of the photons of certain pulse phase interval, in the total number of photons of certain energy range.}
\label{edlc_col}
\end{figure}

\begin{figure*}[t]
\centering
\includegraphics[width=130mm]{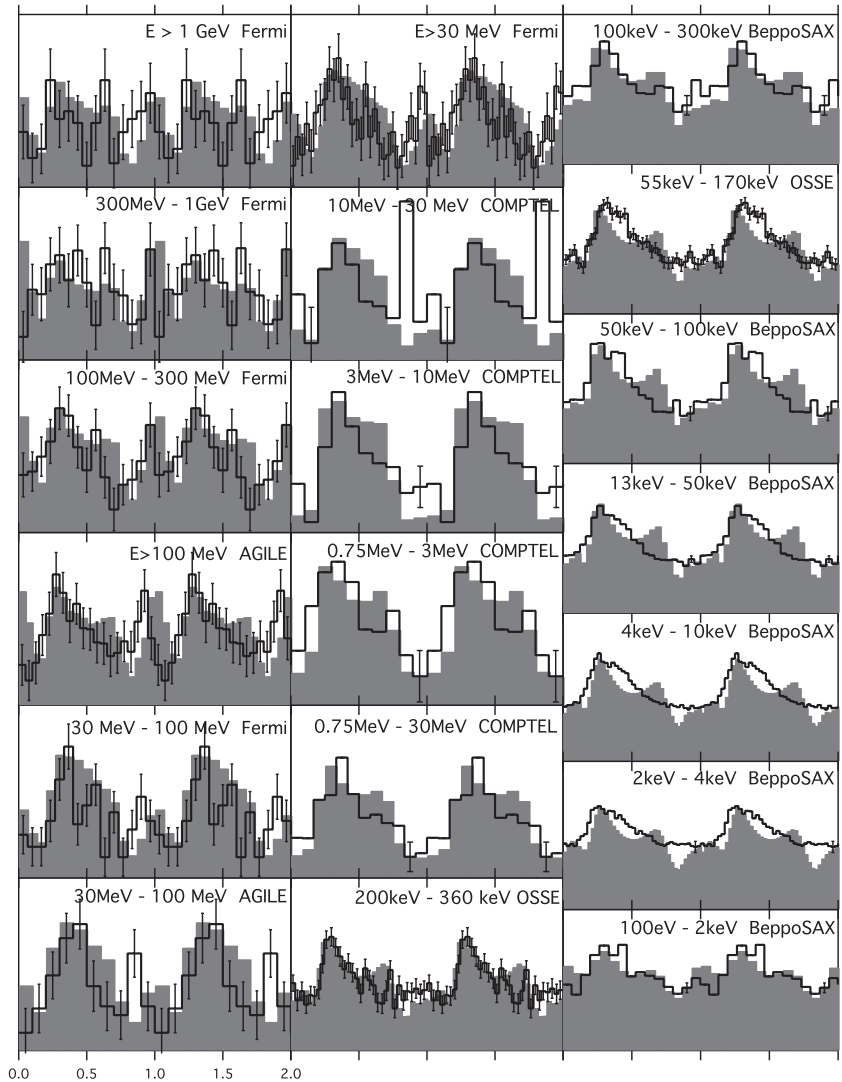}
\caption{The simulated energy dependent light curves (grey histograms), comparing with the observed data (solid lines), 
which are provided by Fermi \cite{Fermi_Obs}, AGILE \cite{AG_Obs}, COMPTEL \cite{Ginga_Obs}, OSSE \cite{OSSE_Obs} and BeppoSAX \cite{Beppo_Obs}.}
\label{edlc}
\end{figure*}

If the radiation satisfies $|\zeta-\beta|<0.5^{\circ}$, where $\beta$ is our viewing angle, we calculate its spectrum. Because some synchrotron photons approach the star and become pairs, we trace each direction of 
 the synchrotron radiation to calculate the attenuation caused by the absorption of the magnetic field. The simulated phase averaged spectrum is the sum of the visible survival synchrotron radiation, which is shown by Figure~\ref{Spectrum}.
 The energy dependent light curves are obtained by integrating the phase resolved spectra. The number of the photons with $E_1\le{}E\le{}E_2$
 measured at pulse phases between $\psi_1$ and $\psi_2$ is calculated from
\begin{equation}
N_{\gamma}(E_1, E_2, \psi_1, \psi_2)\propto\int^{E_2}_{E_1}{F_{tot}(E, \psi_1 \leq \psi \leq \psi_2)}dE.
\label{int_spec}
\end{equation}
By fitting the observed spectrum and the energy dependent light curves, we choose the inclination angle as $\alpha=20^{\circ}$, and the viewing angle as $\beta=11^{\circ}$. 
Figure~\ref{edlc_col} is the simulated energy dependent light curves, and the comparisons between the simulated light curves and the observed ones are given by Figure~\ref{edlc}.

\section{DISCUSSION}
\begin{itemize}
\item{Where are the GeV curvature photons?}

The magnetosphere of PSR B1509 can generate GeV $\gamma$-ray photons by accelerating the charged particles in the outer gap, as well as other $\gamma$-ray spin-down powered pulsars.
However, these outgoing curvature photons are missed by our line of sight (Figure~\ref{lc_colmap}). As shown in Figure~\ref{fig9}, if the viewing angle or inclination angle increases, the GeV curvature photons become visible.

\begin{figure}
\includegraphics[width=70mm]{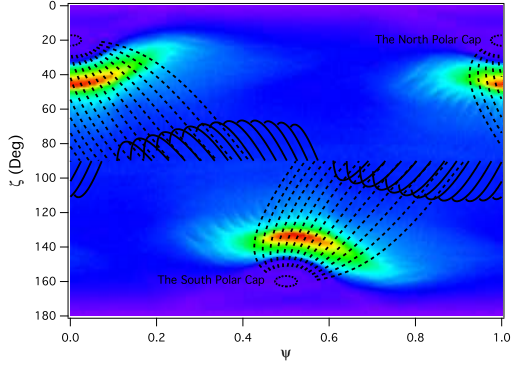}
\caption{The skymap of the viewing angle $\zeta$ and the pulse phase $\psi$ of the outgoing curvature photons (solid line) emitted from the null charge surface to the light cylinder,
  the incoming curvature photons (dashed line) emitted from the null charge surface to the stellar surface, and the synchrotron photons (color) emitted from the places 
  where the incoming curvature photons become pairs. The inclination angle is $\alpha=20^{\circ}$.}
\label{lc_colmap}
\end{figure}

\begin{figure}
\centering
\includegraphics[width=70mm]{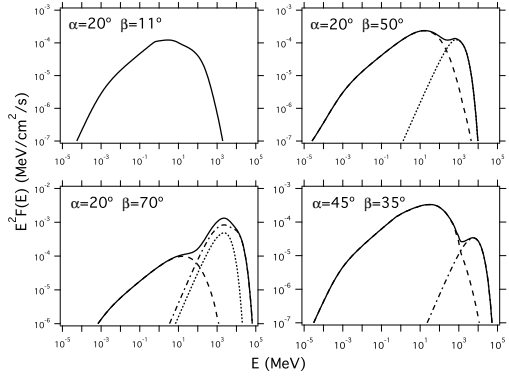}
\caption{The simulated phase averaged spectra of different inclination angles $\alpha$ and different viewing angles $\beta$. 
The dashed lines are the synchrotron radiations of the secondary pairs, the dot-dashed lines are the outgoing curvature radiations, 
the dot lines are the survival incoming curvature radiations, and the solid lines are the total spectra.}
\label{fig9}
\end{figure}

\item{Why is the cut-off energy of sepctrum so low?}

This is due to the absorption of the magnetic field. The magnetic field of PSR B1509 converts not only the curvature photons, which are emitted by the incoming particles, to pairs, 
but also part of the synchrotron photons emitted by these pairs. Figure~\ref{fig5} is the spectra of $\alpha=20^{\circ}$ and $\beta=11^{\circ}$, where the outgoing curvature 
radiation and the possible survival incoming curvature radiation are missed by the line of sight. The dashed and solid lines are the cases with and without the attenuation of the synchrotron 
radiation of the secondary pair caused by the magnetic field. 
 
\begin{figure}
\centering
\includegraphics[width=70mm]{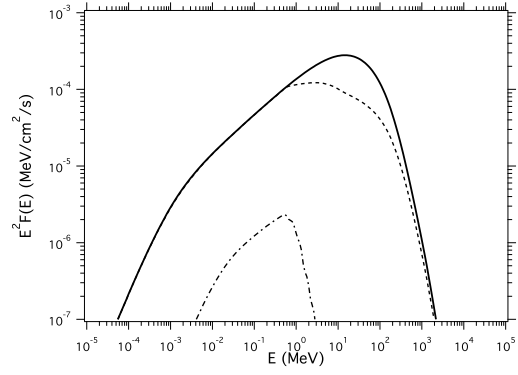}
\caption{The spectra of the synchrotron radiations of the secondary pairs generated by pair creations. The solid line is the one without considering
 the attenuation of the synchrotron photons by the magnetic pair creation. The dashed line is the spectrum of the survival synchrotron photons from the magnetic pair creation. 
 The dot-dashed line is the contribution of the photon-photon pair creation, which considers the attenuation caused by the magnetic pair creation.}
\label{fig5}
\end{figure}

\item{Why is there a second peak at phase$\sim$0?}

The second peak of PSR B1509 is the synchrotron radiation emitted by the pairs with smaller pitch angles, 
which are converted from the curvature photons with higher energy emitted from the region of $20R_s<r<30R_s$ (Figure~\ref{fig10}). 
For a fixed birth place of the curvature photons $\vec{r}$, the photons with higher $E_{cur}$ become pairs earlier. These photons with higher energy can become pairs under smaller pitch angle $\theta_p$,
 and the pairs are generated further from the star. If a synchrotron photon is emitted further from the star, it has higher chance to survive from the absorption of the magnetic field. 
 When $\beta<\alpha$, the pulse phase of the synchrotron radiation is closer to 0 when the pitch angle 
is smaller, and closer to 0.5 when the pitch angle is larger. 

\begin{figure}
\centering
\includegraphics[width=70mm]{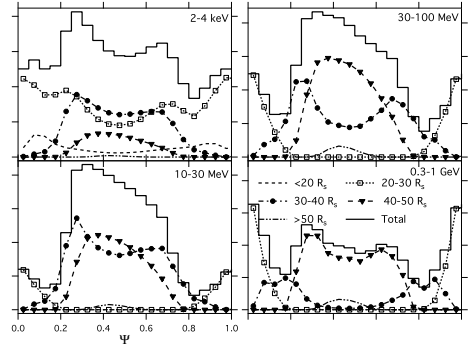}
\caption{The energy dependent light curves, whose original curvature photons are generated in five regions: $r\le{}20R_s$, $20R_s<r\le{}30R_s$, $30R_s<r\le{}40R_s$, $40R_s<r\le{}50R_s$, and $r>50R_s$.}
\label{fig10}
\end{figure}

\item{Why does the second peak show up when energy increases?}

If the particle is made via the magnetic pair creation, the typical energy of its synchrotron radiation is
proportional to the energy of its original curvature photon.
The curvature photon with higher energy can become pair under smaller pitch angle, which makes the pulse phase of the synchrotron radiation close to 0.
Therefore, for the same birth place of the curvature photons, if the viewing angle is smaller than the
inclination angle, the observed spectrum of the synchrotron radiation, whose pulse phase is closer to 0, has higher cut-off energy.

The viewing angle $\beta$ determines the 
energy where the second peak shows up. If $\beta<\alpha$, the second peak shows up at higher energy when the $\beta$ gets closer to $\alpha$, because the observed second peak is 
contributed by the pairs with smaller pitch angles $\theta_p$, which were the curvature photons of higher energy.

\end{itemize}

\section{CONCLUSION}
The line of sight of PSR B1509-58 is in the direction of incoming beam instead of outgoing beam, otherwise a
characteristic power law with exponential cut-off spectrum with cut-off energy around a few GeV should be
observed. In order to avoid seeing the outgoing flux and fit the observed multi-wavelength light curves we need to 
choose inclination angle=20$^{\circ}$ and viewing angle=11$^{\circ}$. The observed spectrum is the synchrotron radiation
emitted by the pairs produced by the magnetic field that converts the major part of the incoming curvature
photons. The energy of the curvature photon and the magnetic field determine the pitch angle of the pair. We find that the differences between the light curves of different energy bands are due to the different
pitch angles of the secondary pairs, and the second peak appearing at E$>$10MeV is the synchrotron radiation emitted by the pairs with smaller pitch angles. 

\bigskip 
\begin{acknowledgments}
We thank Alice K. Harding and W. Hermsen for useful discussion. This work is supported by a GRF grant of Hong Kong Government under 700911P.
\end{acknowledgments}

%




\end{document}